\documentclass[12pt, a4paper, twocolumn]{article}
\usepackage{xurl}
\usepackage[super,comma,sort&compress]{natbib}
\usepackage{abstract}

\usepackage{times}
\usepackage{epsfig}
\usepackage{graphicx}
\usepackage{amsmath}
\usepackage{amssymb}

\usepackage{caption}
\usepackage{subcaption}
\usepackage{booktabs}
\usepackage{authblk}

\usepackage{hyperref}
\hypersetup{colorlinks=true, urlcolor=blue, linkcolor=blue, citecolor=blue}

\title{Dissecting U-net for Seismic Application: An In-Depth Study on Deep Learning Multiple Removal}

\author[1,2]{Ricard Durall$^*$}
\author[1,2]{Ammar Ghanim$^*$}
\author[1,2]{Norman Ettrich}
\author[1,2,3]{Janis Keuper}
\affil[1]{Fraunhofer ITWM}
\affil[2]{Fraunhofer Center Machine Learning}
\affil[3]{IMLA, Offenburg University}

\date{}

\newcommand{\abstractText}{\noindent
Seismic processing often requires suppressing multiples that appear when collecting data.
To tackle these artifacts, practitioners usually rely on Radon transform-based algorithms as post-migration gather conditioning.
However, such traditional approaches are both time-consuming and parameter-dependent, making them fairly complex.
In this work, we present a deep learning-based alternative that provides competitive results, while reducing its usage's complexity, and hence democratizing its applicability.
We observe an excellent performance of our network when inferring complex field data, despite the fact of being solely trained on synthetics.
Furthermore, extensive experiments show that our proposal can preserve the inherent characteristics of the data, avoiding undesired over-smoothed results, while removing the multiples.
Finally, we conduct an in-depth analysis of the model, where we pinpoint the effects of the main hyperparameters with physical events.
To the best of our knowledge, this study pioneers the unboxing of neural networks for the demultiple process, helping the user to gain insights into the inside running of the network.
}

\begin{document}


\twocolumn[
  \begin{@twocolumnfalse}
    \maketitle
    \begin{abstract}
      \abstractText
      \newline
      \newline
    \end{abstract}
  \end{@twocolumnfalse}
]

\def\thefootnote{*}\footnotetext{Equal contribution}

\section{Introduction}
In seismic exploration, geophysicists interpret reflections of acoustic waves to extract information from the subsurface.
These reflections can be classified as primary or multiple.
Primary reflections are those seismic events whose energy has been reflected once, and they are employed to describe the subsurface interfaces.
Multiples, on the contrary, are events whose energy has been reflected more than once, and appear when the signal has not taken a direct path from the source to the receiver, after reflecting on a subsurface boundary.
The presence of multiples in the recorded dataset can trigger erroneous interpretations, since they do not only interfere with the analysis in the post-stack domain, e.g., stratigraphic interpretation, but also with the analysis in the pre-stack domain, e.g., amplitude variation with offset (AVO) inversion.
For this reason, the demultiple process plays a crucial role in any seismic processing workflow.
Multiple-attenuation methods can be classified as predictability- and separation-based. 
Predictability-based methods exploit the repetitive nature of multiples and their inherent connection to primaries.
In general, such methods consist of a prediction-subtraction step, in which predicted multiples are subtracted from the raw data.
\begin{figure*}{h}
\begin{center}
   \includegraphics[width=.8\linewidth]{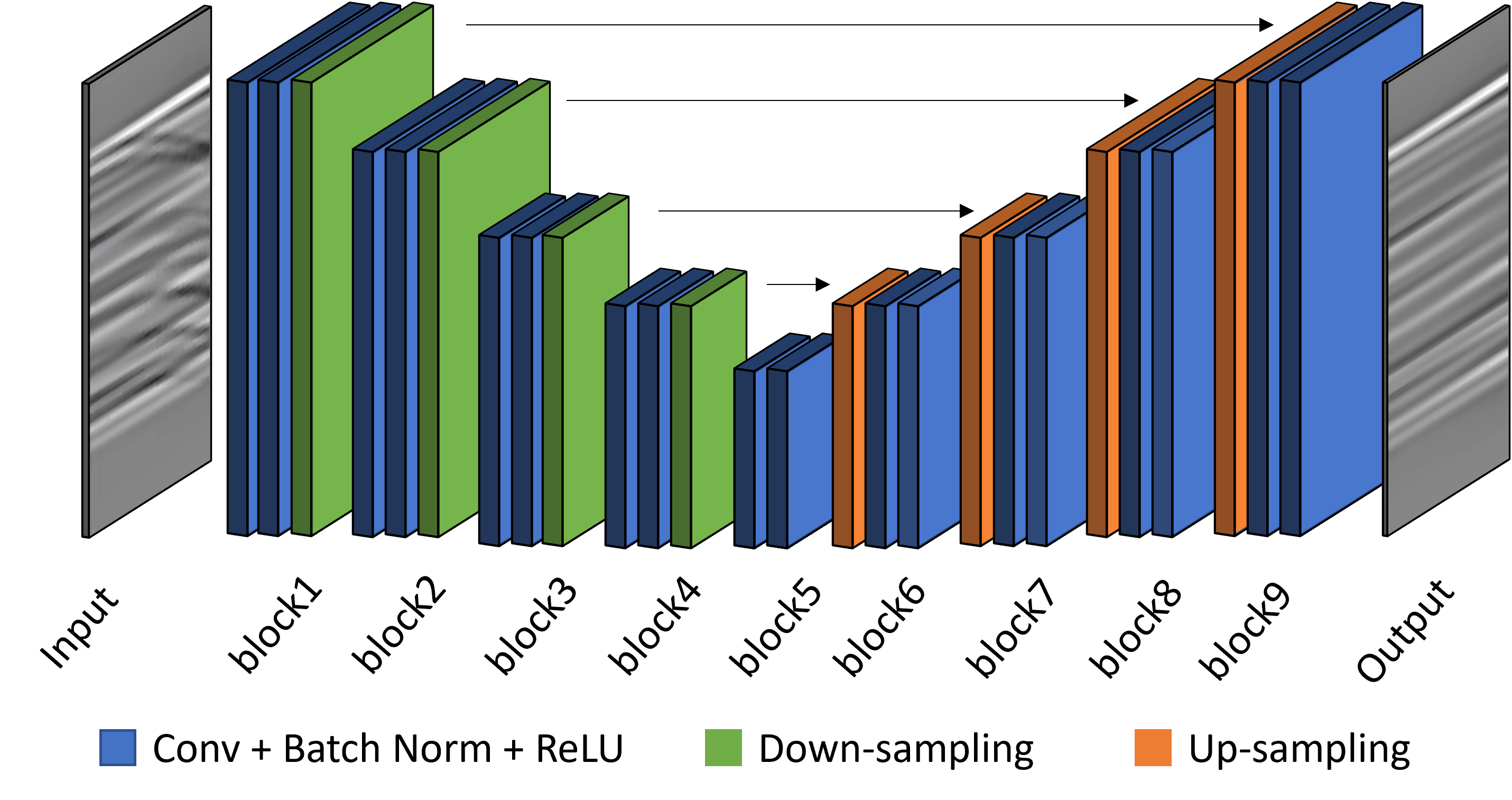}
\end{center}
   \caption{U-net architecture for multiple-attenuation.
   The task of this model is to learn to remove multiples while keeping the rest of the image unmodified, i.e., primaries and data characteristics.}
\label{fig:unet}
\end{figure*}
Some of the most widely used methods are the wavefield extrapolation method \cite{berryhill1986deep,wiggins1988attenuation}, the surface-related multiple elimination (SRME) \cite{berkhout1985seismic,verschuur1991surface,verschuur1992adaptive} as well as the inverse scattering series free-surface multiple elimination (ISS FSME) \cite{carvalho1991examples,weglein1997inverse,weglein2003inverse}.
All these approaches are well-known for their ability to successfully suppress water bottom multiples.
However, these methods are complex and have high computational demand, especially when dealing with 3D structures.
For example, sparse seismic acquisition grids require costly data interpolation to avoid aliasing \cite{gisolf2010principles}.
Moreover, they heavily rely on the quality of near-offset data and consequently, if this data is damaged or misses some traces, the reliability of the models is drastically decreased.
On the other hand, separation-based methods have a simpler nature and are less computationally intensive.
They translate seismic data into intermediate domains, where different characteristics of multiples and primaries are accentuated.
In this way, one can eliminate multiples at will \cite{weglein2011multiple}.
The idea behind this method is to exploit the fact that on average multiples have encountered a lower velocity than the primaries, thus multiples are expected to exhibit an increasing move-out along the offset.
One of the most widespread methods making use of this feature is the parabolic Radon transform (RT) \cite{hampson1986inverse}.
It translates pre-stack gathers from a time-offset to a $tau$-p space, by mapping them by a set of parabolic events, while optimizing their least-squares fit to the data.
When applying RT to remove multiples, it is assumed that residual move-outs of primaries and multiples follow an idealized parabolic pattern, and there is no truncated offset axis.
However, multiples often do not follow a perfect parabolic form on normal move-out (NMO) corrected gathers \cite{hampson1986inverse}, and data certainly does have truncated traces.
As a consequence, RT can potentially degrade parts of the primary signal.
Another limitation appears when dealing with gathers that are coarsely sampled.
In such cases, RT can lead to insufficient separation of primaries and multiples, creating residual multiple energy or aliasing effects in the time-offset domain.
To address some of the aforementioned weaknesses, the high-resolution Radon multiple removal was introduced \cite{sacchi1995high,sacchi1999fast,trad2003latest}.
However, this method requires a higher complexity, resulting in numerous parameters that the interpreter needs to manually fine-tune. 
When it comes to industry workflows, it is typical to combine the usage of predictability-based methods, e.g., SRME, in the pre-migration domain, and a separation-based method, e.g., RT demultiple, in the post-migration.
In this fashion, interpreters can leverage the best from both methodologies, and achieve more reliable outcomes.
With the introduction of deep learning, a new vein of methods has emerged \cite{breuer2020deep,bugge2021demonstrating}.
These approaches are based on artificial neural network architectures, which are universal approximators, i.e., they can, in theory, model any continuous function.
To the best of our knowledge, deep-learning models have focus only on the post-migration domain.
\cite{breuer2020deep} presented a deep learning-based method to trim statics and to remove multiples (using  move-out discriminator process) using synthetic training data.
\cite{bugge2021demonstrating} proposed a similar topology that simultaneously tackled the problem demultiple and denoising.
Although deep-learning approaches have been a breakthrough in the demultiple field, they depend on large amounts of annotated training data, which limits its application.

\begin{figure*}[h]
\begin{center}
   \includegraphics[width=\linewidth]{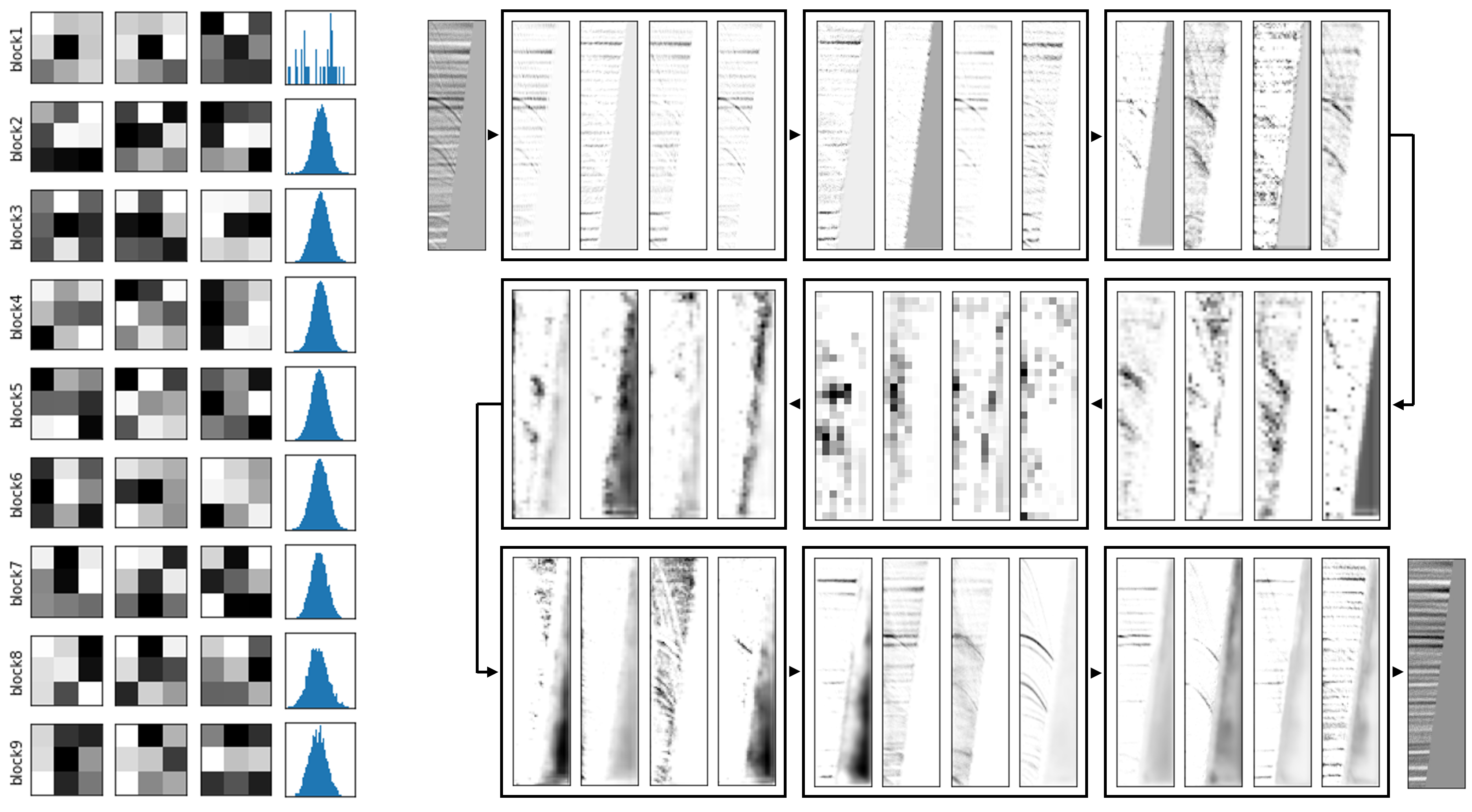}
\end{center}
   \caption{Visualization of U-net inner structure after each block.
   (Left) From top to bottom, each row displays three random filters, and the histogram's statistics from all the filters from each block, where the x-axis is the weight values and the y-axis the frequency of apperance.
   (Right) From upper left to bottom right, following a ``Z'' shape, the transformations that the input image undergoes before the demultiples are removed.
   Each group shows four random feature maps, and belongs to a block of the U-net structure (see Figure \ref{fig:unet}).}
\label{fig:inner}
\end{figure*}

In this paper, we introduce an automated end-to-end deep-learning approach, which successfully removes those events that follow parabolic alike patterns, while preserving the primary energy at crosspoints.
It falls under the separation-based methods, as it deals with NMO corrected gathers, tackling discriminating events with different move-outs.
Furthermore, our approach works in a parameter-free manner, alleviating the user from any manual task.
To achieve that, we train a convolutional neural network with synthetic pairs of multiple-infested and multiple-free gathers.
Besides, we conduct an in-depth hyperparameter search, where we study the role that different components play, and their impact into the outcomes.
To that end, we visualize the inner workings of our neural network, to pinpoint the effect of the main hyperparameters to physical events.
By means of this pioneer study, i.e., the unboxing of neural networks for the demultiple process, we aim to gain insights of the inside running of the network.
Finally, extensive in-field evaluations show that our model is able to preserve the inherent characteristics of the data in different scenarios, and thus, to generalize well.
As a result, our approach can be seen as an alternative to traditional separation approaches in post-migration stage, such as parabolic RT, in existing processing workflows.

\section{Visualization of U-net}
U-net \cite{ronneberger2015u} is a convolutional neural network (CNN) topology, which was initially designed for semantic segmentation tasks in the medical domain.
However, due to its generalization capacity, it has been widely adapted to many domains.
U-net's architecture is split into two paths: the contraction path, also called encoder, which is used to capture the context in the image; and the expanding path, also called decoder, which is used to enable precise localization.
Both paths are symmetric and made of blocks of convolutional layers followed either by a down-sampling operation (encoder), or by an up-sampling operation (decoder).
In addition to the encoder-decoder scheme, U-net has long skip connections that bypass some layers, and connect different blocks from the encoder to their counterparts from the decoder. 
These shortcuts provide alternative paths for the gradient during back-propagation, that help the model to incorporate fine-grained details in the predictions.
Figure \ref{fig:unet} shows the architecture of U-net for the demultiple scenario.

CNN architectures are successfully used in a large variety of applications, ranging from computer vision to natural language processing.
They are made up of neurons that have learnable parameters arranged in filter-shape structures.
Each of these neurons receives some inputs, performs a dot product and finally, applies a non-linear activation function. e.g., sigmoid or rectified linear unit (ReLU) \cite{nair2010rectified}.
The output of the activations for a given filter is called feature map, or activation map.

Although the learning mechanism (back-propagation) is well understood, the intrinsic details, such as the reason why a specific decision or prediction is made, are not.
As a result, neural networks are typically treated as black box models.
To better understand the internal working, we visualize different components of the network.
In particular, we investigate the filters and the feature maps to try to conceptually unravel the learning of the model, when dealing with demultiple problems.

On the left side of Figure \ref{fig:inner}, we can see some filters that the network has learnt.
Seemingly, they do not seem to display any human-recognizable pattern from which one can draw conclusions.
More informative is, however, the statistics.
The filters' weights appear to always follow a Gaussian distribution, independent of the layer.
Similar observations by \cite{gavrikov2022empirical} suggest that convolution filters do not have distribution shifts along different axes of meta-parameters, like data type, task, architecture, or layer depth. 
Nonetheless, we notice that the first block might break this empirical deductions, meaning that depth could indeed play a certain role in shallow layers.
On the right side of Figure \ref{fig:inner}, we can observe some feature maps from different blocks.
These intermediate representations display how the network modifies the input image, and help us to understand how multiples are identified and suppressed.
On the one hand, as expected, we can visually assess a gradual loss of resolution (high-frequency components) in the first blocks, due to their down-sampling operations from the contraction path.
The opposite effect can be seen in the last blocks, caused by the up-sampling operations from the expanding path.
On the other hand, contrary to what might be intuitive, the network is not learning to suppress multiples right from the beginning.
In fact, they are present in all blocks, and almost in all feature maps.
What the network seems to learn, instead, is to identify the multiples in each block to have a fully understanding of the event.
In this manner, in the very last layer, the model combines the feature maps in such a way that the undesirable events (multiples) are cancelled out.

\section{Analysis of U-net Parametrization}
Hyperparameters are values that control the learning process of neural networks.
They define different aspects of the model, such as learning rate, optimizer, depth, activation function, loss function, just to mention a few.
In general, neural networks are notorious for being very sensitive to the choice of hyperparameters, resulting in fairly different outcomes when the parameters are slightly modified.

In this section, we identify and describe the empirical effects that some hyperparameters have on our multiple-attenuation network.
In particular, we focus on the impact of the optimizer, sampling technique, kernel size, loss function and depth.
To that end, we average validation results of five independent runs to guarantee reproducibility.
We evaluate these results on four different metrics: mean square error (MSE), signal-to-noise ratio (SNR), structural similarity (SSIM) and peak correlation (PCORR).
Furthermore, we validate the outcome on synthetic and real datasets.
In this manner, we ensure certain generalizability and neutrality in our observations.

\begin{figure*}[p]
  \begin{subfigure}{1\textwidth}
  \centering
   \includegraphics[width=\linewidth]{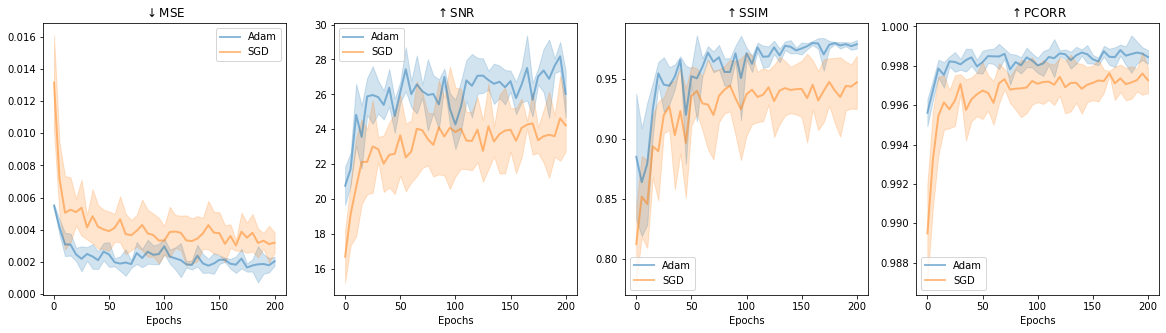}
   \caption{Optimizer assessment based on different quantity metrics.}
\label{fig:optimizer_metrics}
  \end{subfigure}\\[.5ex]
  \begin{subfigure}{1\textwidth}
  \centering
   \includegraphics[width=\linewidth]{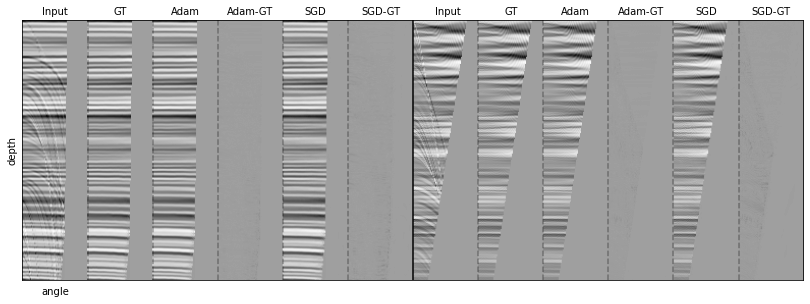}
   \caption{This figure displays two random gathers from our validation synthetic dataset.
   We infer results on a pretrained model with Adam and on a pretrained model with SGD.
   Both outcomes are multiple-free, and their differences with the ground-truth (GT) are neglectable.}
\label{fig:optimizer_synthetic}
  \end{subfigure}\\[.5ex]
    \begin{subfigure}{1\textwidth}
  \centering
   \includegraphics[width=\linewidth]{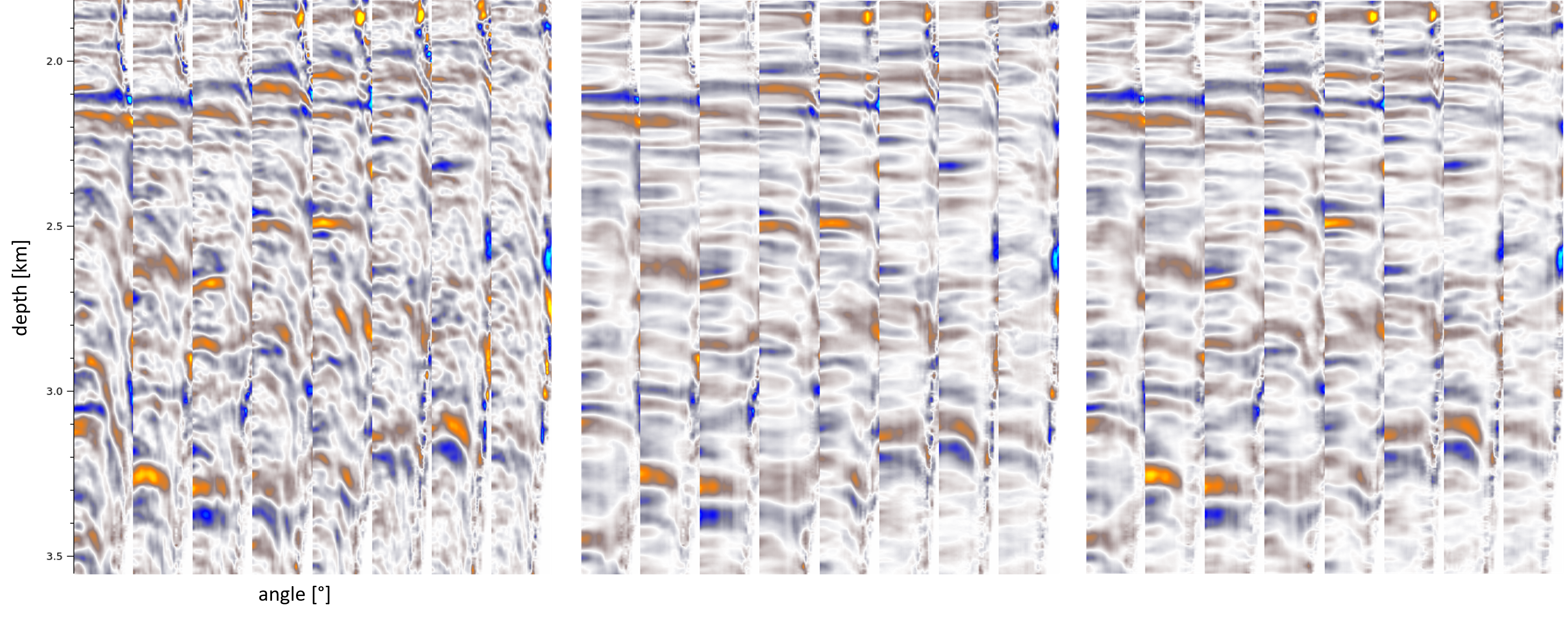}
   \caption{This figure illustrates a collection of eight gathers from real data.
   From left to right, the input data (with multiples), the output from a pretrained model with Adam, and the output from a pretrained model with SGD.}
\label{fig:optimizer_real}
  \end{subfigure}\\[1ex]
  \caption{The optimizers employed in the neural network might converge into different local minimas.
  Therefore, we study the impact of two approaches: SGD and Adam.
  The figures above show how these optimizers behave under synthetic as well as real scenarios.}
  \label{fig:optimizer}
\end{figure*}

\subsection{Optimization Functions}
Optimizers can be explained as a mathematical algorithm that modifies the weights of the network to minimize the loss function.
They are built upon the idea of gradient descent, i.e., the greedy approach of iteratively decreasing the loss function by following the gradient.
There are two main groups of optimizers: adaptive and non-adaptive methods.
\cite{hardt2016train} argued that non-adaptive methods, such as stochastic gradient descent (SGD), are conceptually more stable for convex and continuous optimization, having smaller generalization error.
They also proved that, under certain conditions, the results can be carried over to non-convex loss functions.
Follow-up work\cite{wilson2017marginal}, found empirical evidence of the poor generalization performance of adaptive optimization methods, such as adaptive moment estimation \cite{kingma2014adam} (Adam).
Even when the adaptive methods achieve the same training loss or lower than non-adaptive methods, the test performance is worse.
Finally, \cite{choi2019empirical} claimed that the hyperparameter of the optimizer could be the reason that adaptive optimization algorithms failed to generalize.

In our experiments, we evaluate the impact of SGD with momentum and Adam optimizers for the demultiple task.
Figure \ref{fig:optimizer_metrics} shows the validation metrics in synthetic data for the two selected optimizers.
In these plots, we can observe how the adaptive optimization (Adam) converges faster than the non-adaptive one (SGD), and also ends up in lower local minima, i.e., all the metrics reach better values.
Nonetheless, although the gap between both optimizers seems to be significant, when inspecting synthetic results, the differences are negligible (see Figure \ref{fig:optimizer_synthetic}).
Furthermore, surprisingly, the demultiple outcomes on real dataset suggest that the model trained with Adam optimizer tends to fail to generalize more often, and its results are not always consistent, varying among different runs.
In Figure \ref{fig:optimizer_real} we display some results on real data, where  we see how Adam's approach suppresses occasionally the primary energy, as for the reflection at 3.25km in the second gather from the left, and leaves some residual multiples in the far-stack, as for the reflection at 2.1km in the sixth gather from the left.
Despite that our model is trained using synthetic data, the system is meant to be applied on real data.
Therefore, we favour to employ the SGD optimizer.
\begin{figure*}[h]
  \begin{subfigure}{1\textwidth}
  \centering
   \includegraphics[width=.3\linewidth]{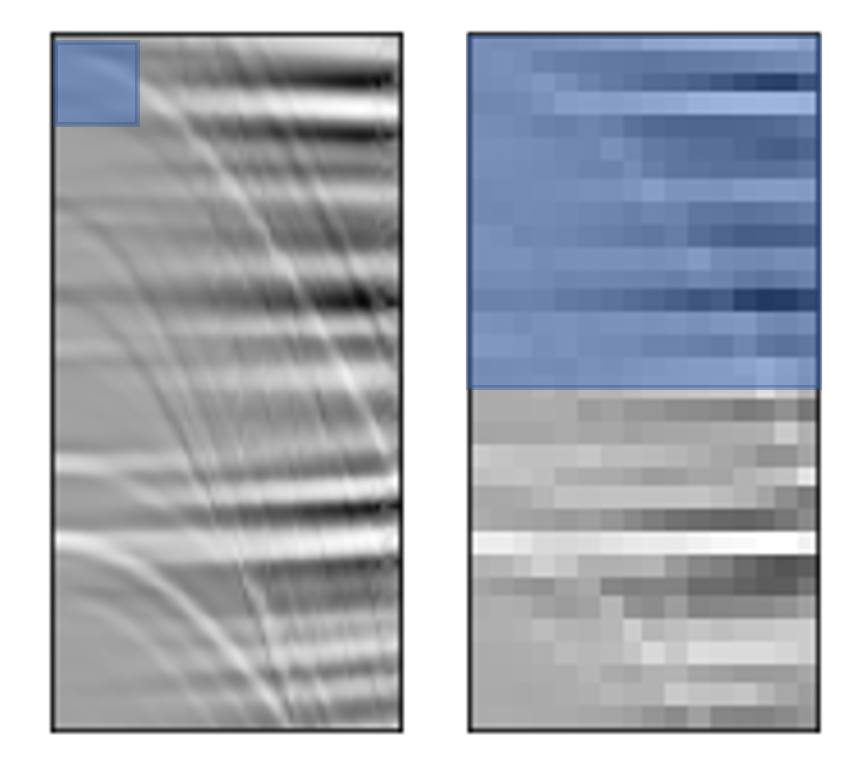}
   \caption{This figure illustrates the effect of down-sampling an image.
   (Left) Original image, 64 $\times$ 128 pixels.
   (Right) Down-sampled image, 16 $\times$ 32 pixels.
   Given that the kernel size is fixed, in this case 16 $\times$ 16, we can observe how the receptive filed in the down-sampled image reaches to cover locations that cannot be covered in the original image, e.g., half of a multiple.
   Note that for our experiments, the kernel size is fixed to 3 $\times$ 3. }
\label{fig:receptive}
  \end{subfigure}\\[1ex]
  \begin{subfigure}{1\textwidth}
  \centering
   \includegraphics[width=\linewidth]{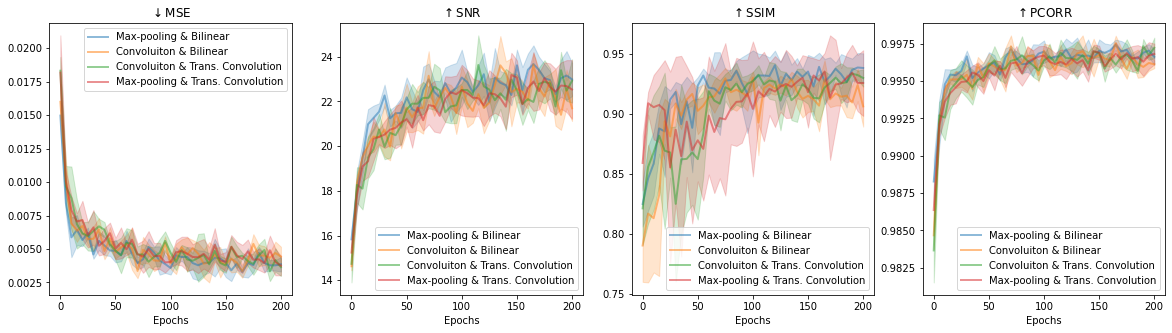}
   \caption{Sampling assessment based on different quantity metrics.}
\label{fig:sampling_metric}
  \end{subfigure}
  \caption{The sampling technique determines how the system modifies the scale of the input images.
  Depending on the task, it can lead to undesired artifacts \cite{durall2020watch}.
  For this reason, we evaluate the effect on our seismic scenario.}
  \label{fig:sampling}
\end{figure*}

\subsection{Sampling Technique and Kernel Size}
CNN-based models gradually down-sample their inputs so that the receptive fields of the deeper filters can reach most of the image at a certain depth.
By doing that, the pixel dependencies, that lie far away from each other in an image, can be captured (see Figure \ref{fig:receptive}).
This is an important aspect for any neural network that needs to interact with content that is spread on the input image, such as in fault detection and in multiple removal.
In our study, we conduct a two-fold analysis related to the sampling; we evaluate the effect of different sampling techniques, and we analyse the impact of the kernel size when sampling.

Sampling techniques refer to those methods that decrease or increase the size of an input.
In the contraction path of U-net, there are two down-sampling approaches: pooling operation and convolution operation.
While the pooling operation does not have learnable parameters (less computational demanding), the convolutional operation does have such parameters.
As a consequence, the latter can capture additional information, whereas the pooling will always imply a loss of information.
In the expanding path, the decoder recombines the features sequentially until it recovers the original input size.
To that end, this path requires up-sampling operations.
Similarly to the contraction path, there are two main approaches: interpolation operation and transposed convolution operation.
The first type of operation is parameter-free and lossy, and the second the opposite.
In order to evaluate the impact of the sampling methods, both down- and up-sampling, we check the different combinations.
For the sake of simplicity, we restrict our analysis to the default configuration which are max-pooling as a non-learnable down-sampling technique, and bilinear as a non-learnable up-sampling technique.

Based on Figure \ref{fig:sampling_metric}, experiments with transposed convolutions have less stable runs, nonetheless, all the sampling techniques reach similar performance.
Therefore, the extra computational cost of the learnable operations is not justified.
Furthermore, the combination max-pooling \& bilinear, being both non-learnable sampling methods, provide the most stable results.
Testing with synthetic and real data show no difference among the configurations.

\begin{figure*}[p]
  \begin{subfigure}{1\textwidth}
  \centering
   \includegraphics[width=\linewidth]{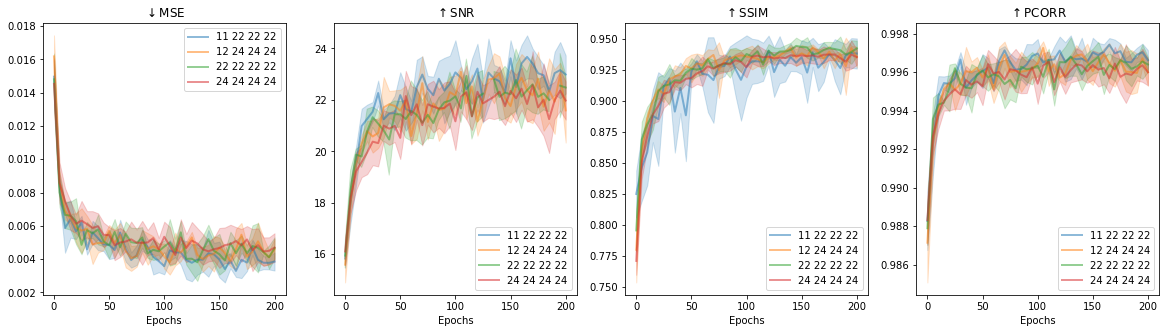}
   \caption{Kernel assessment based on different quantity metrics.}
\label{fig:kernel_metric}
  \end{subfigure}\\[.5ex]
    \begin{subfigure}{1\textwidth}
  \centering
   \includegraphics[width=\linewidth]{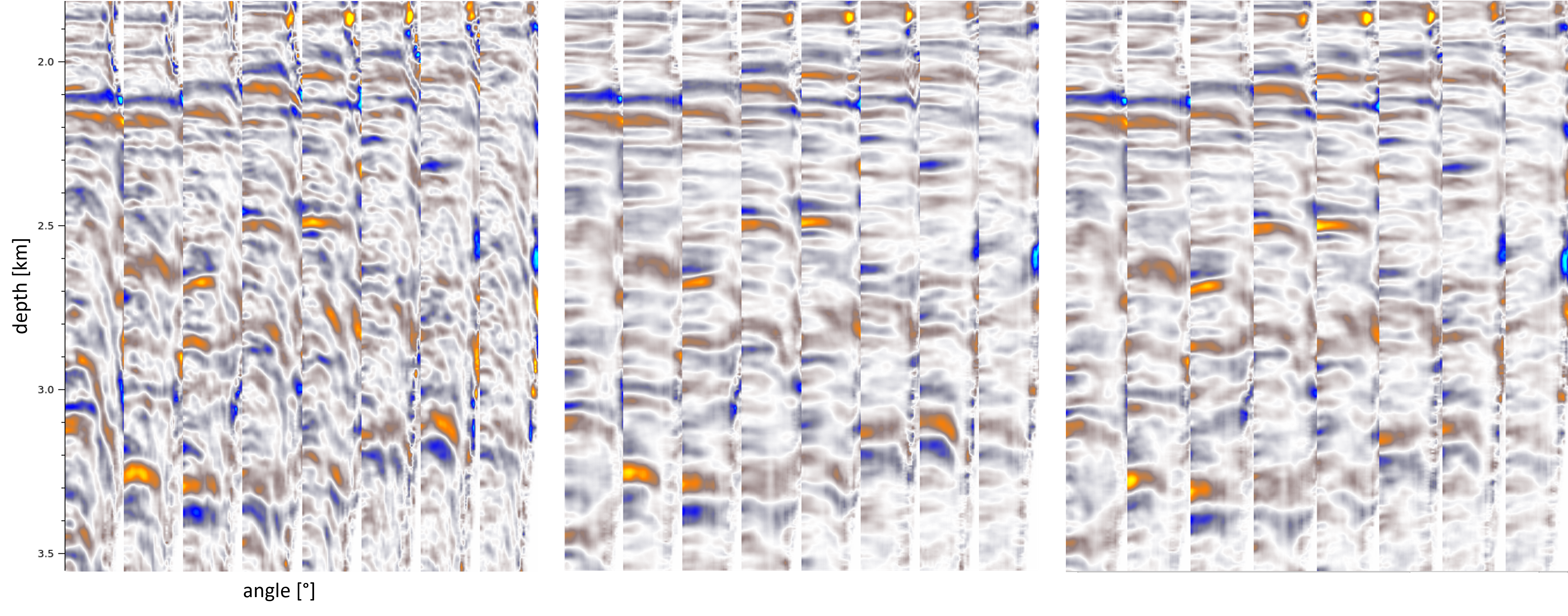}
   \caption{This figure illustrates a collection of eight gathers from real data.
   From left to right, the input data (with multiples), the output from case A, and the output from case D.}
\label{fig:kernel_real}
  \end{subfigure}\\[1ex]
  \caption{ Intuitively, the larger kernel size, the better to capture events that lay far apart.
  Nevertheless, the inferring results suggest that those configurations with larger kernels not only over-smooth the results, but also seem to be produce more unstable results.
  The figures above show results for different kernel configurations.}
  \label{fig:kernel}
\end{figure*}

Besides the sampling techniques, the kernel size might also contribute to the final outcomes.
This hyperparameter determines to what degree the sampling operation down- and up-samples the corresponding input.
Given that we work with elongated events, we empirically analyse the impact of kernel with square and non-square shapes, and assess the impact of more aggressive sampling, i.e., the down- and up-sampling factor.
Table \ref{tab:kernel} describes the scenarios of our study cases, and Figure \ref{fig:kernel} displays the inference results.
Although, the validation metrics seem to report the same behaviour for all the kernels (see Figure \ref{fig:kernel_metric}), we observe a consistent improvement after quality control when using 11 22 22 22 kernel sequence (see Figure \ref{fig:kernel_real}).
Models trained with the larger max-pooling kernels appear to remove multiples more aggressively, i.e., over-smoothing results.
Moreover, the models trained with larger kernels seem to be more sensitive to the initial weights than their counterparts trained with smaller max-pooling kernels.

\begin{table*}[t]
\centering
\begin{tabular}{cccc}
\hline
& Kernel shape & Aggressiveness  & Configuration\\
\hline
Case A & Square & Low & 11 22 22 22\\
Case B & Non-Square & High & 12 24 24 24\\
Case C & Square & Low & 22 22 22 22\\
Case D & Non-Square & High & 24 24 24 24\\
\hline
\end{tabular}
\caption{Each case belongs to a particular arrangement of the kernels.
For example, case B (12 24 24 24): the kernel of block 1 is defined as 12, and the following three blocks as 24.
This means that the first block will down-sample its input only along the y dimension by a factor of 2, and the x dimension will remain unmodified.
Then, the second block will down-sample its input along the x dimension by a factor of 2, and by a factor of 4 in its y dimension.
Note that all these operations will be reversed in the expanding part.}
\label{tab:kernel}
\end{table*}

\subsection{Loss Function}
The selection of a loss function is a challenging task that has a direct impact on the model's behaviours.
For this reason, it is important to choose a function that captures the relevant information that needs to be propagated through the network.
In this work, we advocate for the use of MSE for its simplicity and capacity to deal with outliers.
This loss calculates the difference between the model’s predictions $\hat{y}$ and the ground truth $y$, squares and averages it, across the whole dataset ($N$ samples).
Mathematically, it can be formulated as:
\begin{align}
	MSE = \dfrac{1}{2} \sum_{i=1}^{N} (y_i - \hat{y}_i)^{2}
\end{align}

\begin{figure*}[p]
  \begin{subfigure}{1\textwidth}
  \centering
   \includegraphics[width=.4\linewidth]{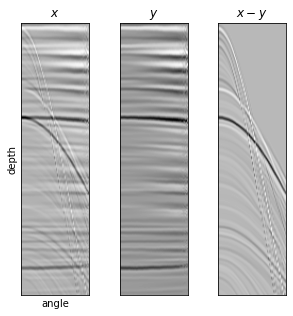}
   \caption{Given an input image with multiples $x$, our goal is to build a network that can eliminate them.
   To achieve that, either we design a model that removes directly the multiples, i.e., targeting $y$; or we design a model that only keeps the multiples, i.e., targeting $x-y$, and then we subtract this result from the input $x$. }
\label{fig:loss_example}
  \end{subfigure}\\[1ex]
  \begin{subfigure}{1\textwidth}
  \centering
   \includegraphics[width=\linewidth]{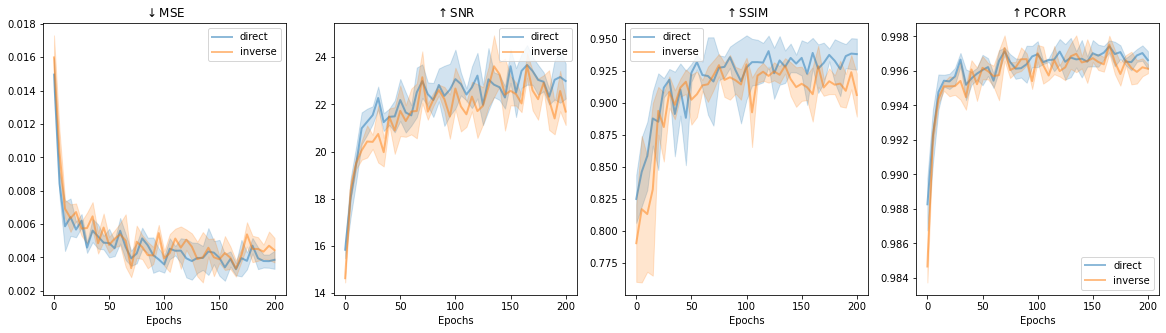}
   \caption{Loss function assessment based on different quantity metrics.}
\label{fig:loss_metric}
  \end{subfigure}
  \caption{These figures describe the two objective functions that we implement in our experiments. }
  \label{fig:loss}
\end{figure*}

In addition to the loss function, we need to define the objective.
In other words, which task the network is targeting.
Here, we propose two objectives: direct and inverse.
Given an input image $x$, the direct proposal tackles the demultiple problem by optimizing the prediction $\hat{y}$, which is a multiple-free image.
The inverse approach, however, formulates the solution from another perspective.
It defines the objective task as an optimization problem, where the prediction $\hat{y}$ should contain only the multiples of the input image, i.e, $x-y$ (see Figure \ref{fig:loss_example}).
In this fashion, the network should focus exclusively on identifying the multiples, omitting the rest.
Once, the model is able to do that, we can simply subtract the prediction from the input image, to obtain a multiple free image.
In Figure \ref{fig:loss_metric}, we have the plots of the metrics using different objectives.
Interestingly, the results from both scenarios are similar.
We hypothesize that the network learns to cancel out the same features, in the direct and inverse formulation, and consequently, the outcomes seem equivalent.
Nonetheless, more advanced loss functions could potentially boost results.

\subsection{Depth of the Network}
The goal of our neural network is to model a function $F$ that maps the raw input data $x$ to a free multiple output.
To that end, we create $F$ by concatenating $n$ non-linear functions $f$, i.e., $F(x) = f_1(f_2(..f_n(x))))$.
Notice that adding more layers provides higher capacity to the network, which leads to deeper networks.
In our experiments, we investigate the effect of three levels of depth.
We take as a baseline the standard model described in Figure \ref{fig:unet}, which consists of nine blocks.
Then, we remove two down-sampling and two up-sampling layers to create a smaller version, called ``small U-net''.
Finally, we repeat the procedure, but this time adding two down-sampling and two up-sampling layers into the baseline.
We call this last model ``big U-net''.
Table \ref{tab:depth} shows the details of each topology, and their inference times.

\begin{table*}[t]
\centering
\begin{tabular}{cccc}
\hline
& \# of block layer & \# of parameters (M)  & inference time (s)\\
\hline
standard U-net & 9 &  17.2 & 22.91 $\pm$ 0.02 \\big U-net & 13 & 276.8 & 42.70 $\pm$ 0.01 \\
small U-net & 5 & 1.0 & 13.67 $\pm$ 0.04 \\
\hline
\end{tabular}
\caption{Adding more capacity might lead to better results, but it also requires higher computational costs.
The last column shows the inference time for each model, when testing on 6000 images 64 $\times$ 256 pixels.
}
\label{tab:depth}
\end{table*}

\begin{figure*}[p]
  \begin{subfigure}{1\textwidth}
  \centering
   \includegraphics[width=\linewidth]{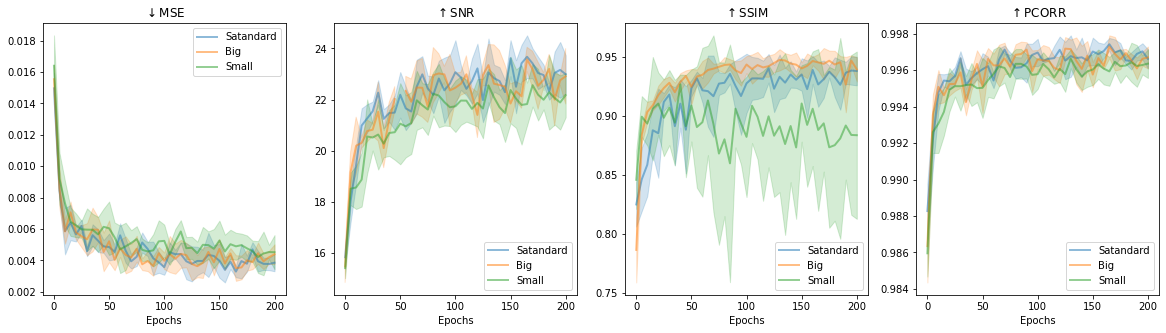}
   \caption{Depth assessment based on different quantity metrics.}
\label{fig:depth_metric}
  \end{subfigure}\\[.5ex]
  \begin{subfigure}{1\textwidth}
  \centering
   \includegraphics[width=\linewidth]{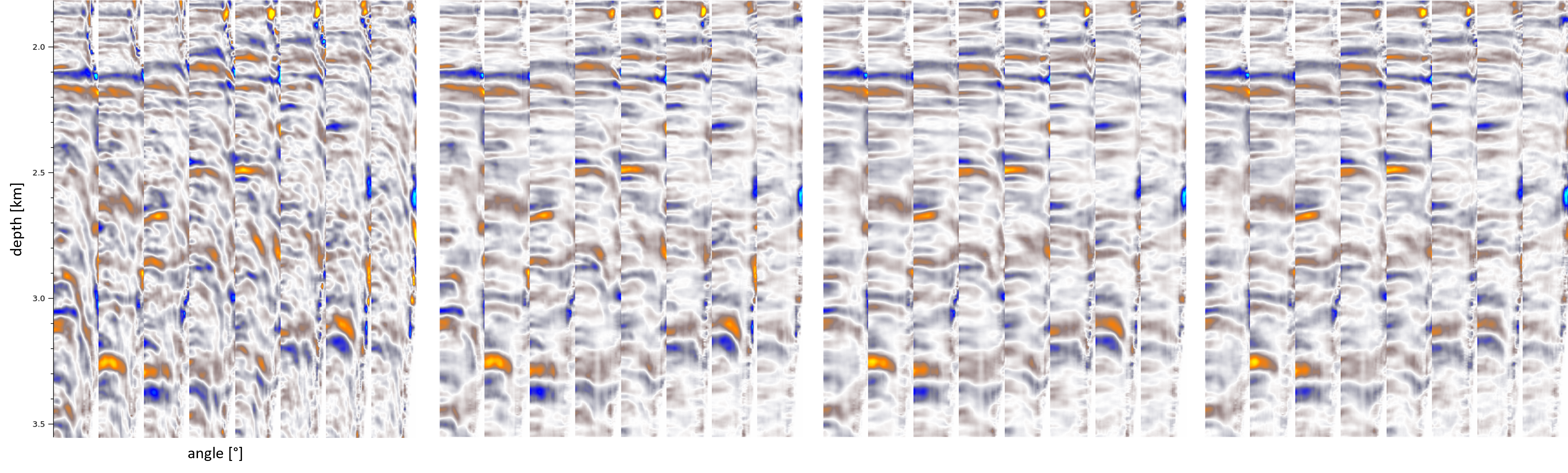}
   \caption{This figure illustrates a collection of 8 gathers from real data.
   From left to right, the input data (with multiples), the output from small U-net, the output from standard U-net, and the output from big U-net.}
\label{fig:depth_real}
  \end{subfigure}\\[1ex]
  \caption{The capacity of a network plays an important role in any learning task.
  Too shallow topologies cannot capture the complexity of the data; too depth can overfit, not improving the final results.
  Therefore, we study the sweet spot for our demultiple task.
  The figures above show how three architectures with different depth behave under synthetic as well as real scenarios.}
  \label{fig:depth}
\end{figure*}

Figure \ref{fig:depth} shows the depth analysis from which we derive the following statements:
(1) The small U-net is too shallow, and does not have enough capacity to suppress the multiples, without over-smoothing the gathers.
As a result, metrics and real data underperform the standard model.
(2) The big U-net model is over-parametrized, and therefore, the extra layers do not offer any further improvement.
In summary, our standard model has the optimal trade-off between quality and size.

\section{Training Dataset}
When interpreting real seismic data, we do not have ground-truth (annotated data).
Unfortunately, this labelled data is one of the cornerstone of any supervised deep learning model.
Manual interpretation is an effective way to acquire ground-truth, but it is an expensive and time-consuming process.
Furthermore, its outcomes rarely contain all the events that would define the characteristics of the subsurface.
To address this issue, in the demultiple scenario, one could create real labelled data, by employing a traditional approach lie RT.
Nevertheless, the network would be biased and limited by the performance of the traditional approach.

\begin{figure}
\begin{center}
   \includegraphics[width=\linewidth]{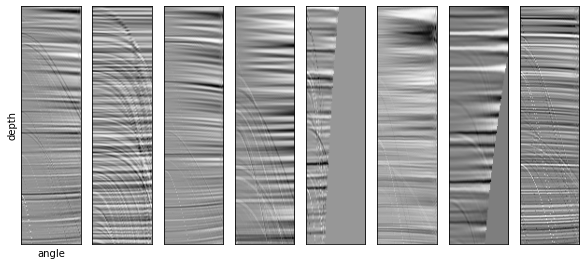}
\end{center}
   \caption{Collection of eight random synthetic gathers used to train.
   Notice that we also generate their multiple free version.}
\label{fig:synthetics}
\end{figure}

In this work, we introduce a network which is able to suppress multiples regardless of the domain and nature of the seismic gathers, i.e., offset or angle domain, and time or depth domain.
To achieve it, we systematically generate synthetic pairs of multiple-infested and multiple-free gathers, thanks to a fine-grained control of this synthetic multiples via a parameter space, which allows us to train models that can generalize.
This parameter space consists of (1) variations of the packing density of multiples and primaries, and their position along the vertical axis; (2) variations of the strength of the RMO effect allowing minimum multiple move-out; and (3) variations of the spectral components of the source wavelets, such as polarity, phase shift, central frequency and bandwidth, together with a central frequency decay along the vertical axis.
Finally, to teach the network to preserve primary energy while removing multiples, we generate synthetic multiples that cross primaries and other multiples (see Figure \ref{fig:synthetics}).

\begin{figure*}[p]
\begin{center}
   \includegraphics[width=\linewidth]{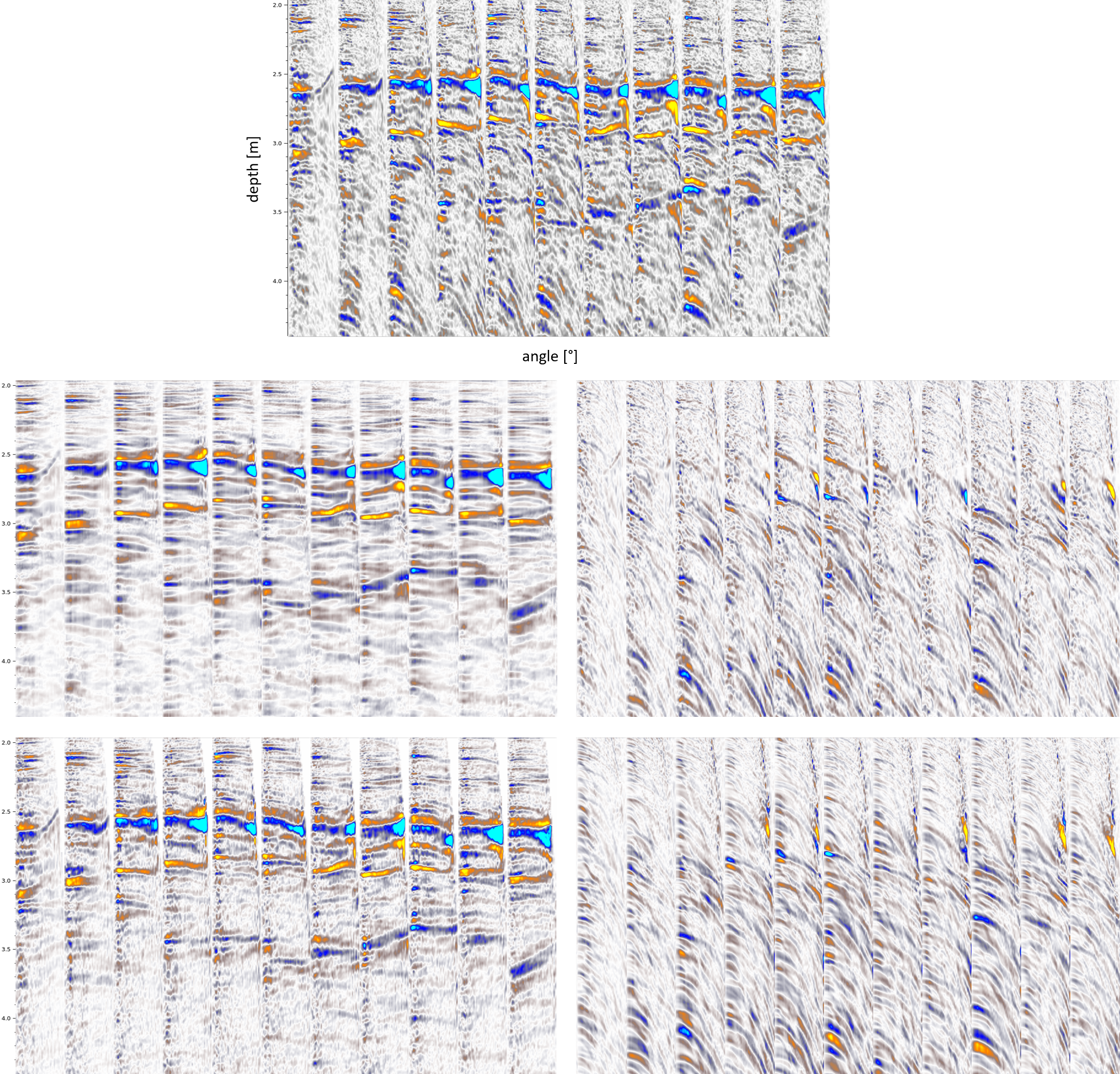}
\end{center}
   \caption{These figures display several pre-stack angle gathers of the Volve dataset.
   First row: Migrated raw angle gathers.
   Second row: Angle gathers, U-net demultiple result (left) and removed multiples (right).
   Third row: Angle gathers, Radon-based demultiple (left) and removed multiples (right).}
\label{fig:prestack_real}
\end{figure*}

\begin{figure*}[p]
\begin{center}
   \includegraphics[width=.9\linewidth]{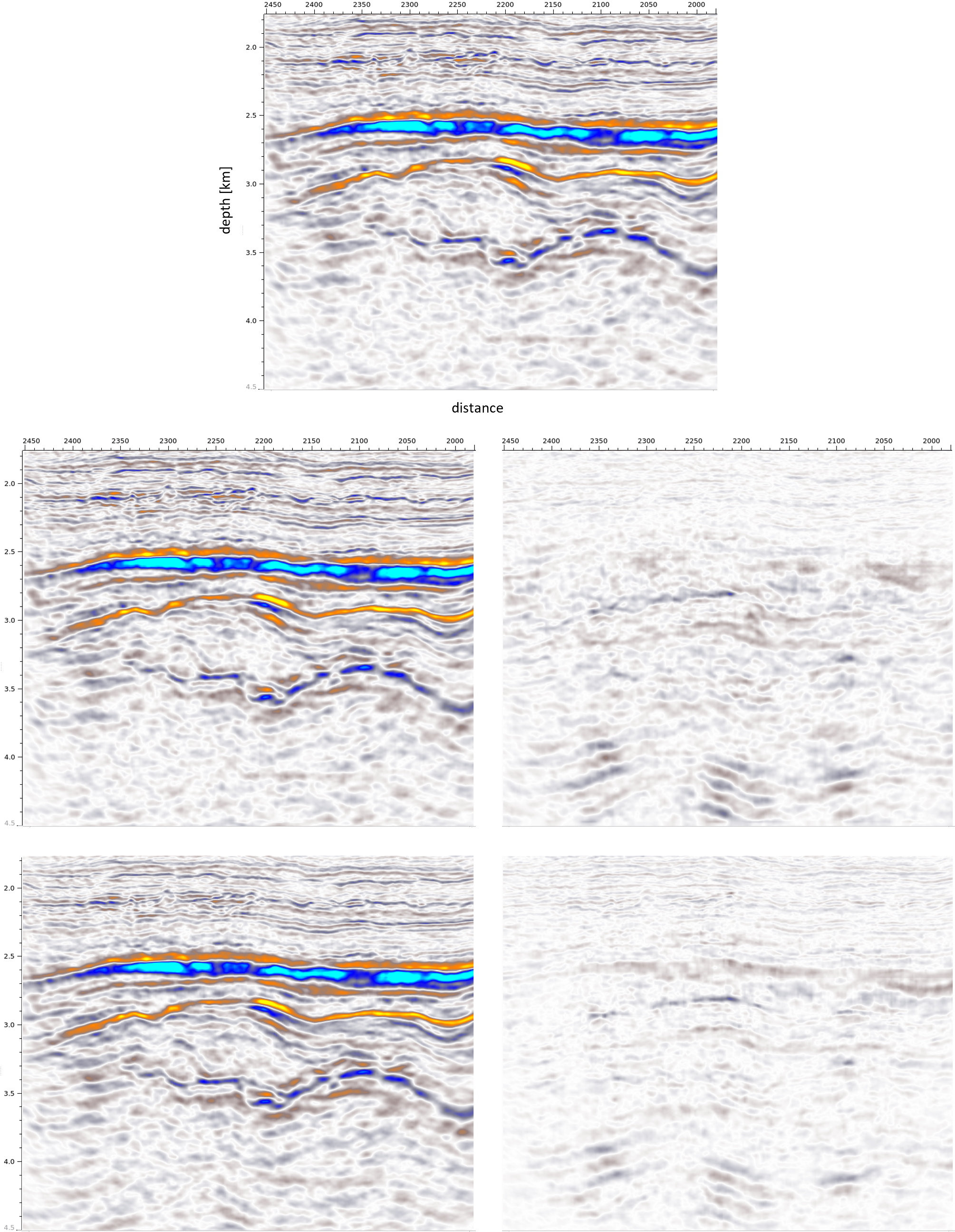}
\end{center}
   \caption{These figures display the full-stack of depth angle gathers of the Volve dataset.
   First row: Migrated raw full-stack section.
   Second row: Full-stack, U-net demultiple (left) and removed multiples (right).
   Third row: Full-stack Radon-based demultiple (left) and removed multiples (right).}
\label{fig:stack_real}
\end{figure*}

The gathers are created by generation of pre-stack reflectivity series by randomly sampling the parameter space.
These are in turn convolved with a wavelet generated by randomly sampling the central frequency and bandwidth parameter space.
These pre-stack gathers are then NMO corrected using a slightly perturbed velocity model; this step allows for introducing the NMO stretch and RMO effects to primaries and multiples.
This, together the definition of a minimum of allowed move-out for the removed multiples, teaches the network not to remove potential non-flattened primaries.

\section{Field Examples}
The approach has been widely tested on real data.
Figure \ref{fig:prestack_real} shows the results of our method as compared to a traditional Radon-based demultiple approach.
Furthermore, we plot the removed multiples for both methods to help to visualize the main discrepancies between both system.
From such a visualization, we observe that the traditional method remove only multiples that follow idealized parabolas, which is highly unlikely for a real dataset, as the multiples, just like the primaries, travel through the overburden and suffer transmission effects.
On the other hand, the events removed by our deep learning method do not always follow an overidealized parabolic course, and they seem more realistic for a dataset acquired in a complex geologic setting.
Moreover, our U-net proposal is better at removing events, which are most likely residuals of a demultiple process in the pre-migration step, i.e., which appear only in the far stack.
Such events can be seen in the far-stack of the first three gathers in \ref{fig:prestack_real}.
We also provide in Figure \ref{fig:stack_real} the results of the same dataset as a full-stack section.
Herein, we can see how the lateral coherency of the removed events is consistent in both approaches, however, the U-net model appears to better follow the overlaying structure and leading to sharper results. 

\section{Discussion}

In this work, we propose a demultiple model that can be interpreted as an image-to-image transformation system.
Given an input data, our deep learning approach identifies the multiples and cancels them out from the output result.
The main success of our implementation is not only the ability to remove multiples, but to do it while preserving the high-frequency components that characterize the data.
Although denoising is common post-processing linked to these frequency components, a non-controlled application of it can lead to loss of nervosity of the data (over-smoothed results), resulting in a decrease of relevant features.
We believe that seismic interpretation is per se a challenging task, therefore, any processing method would need to guarantee the preservation of the characteristics, and a full control of its outputs.

In general, it is relatively trivial to train a neural network that can yield accurate results on a synthetic dataset.
However, it is highly challenging to obtain similar performance on unseen real data with potentially very different acquisition, geology and processing settings.
For this reason, producing synthetic data that mimics the underground events is a crucial ongoing research \cite{durall2021generative}.
During our experimental evaluation, we have iteratively modelled different synthetic data, investigating the effects on real data.
This empirical methodology has allowed us to generate a concise multiple-oriented dataset, with high generalization properties.
In this way, our approach can be used to speed up interpretation tasks, easing human experts to deal with huge volume of real data.

Finally, we think that the geoscience community would benefit from an attempt to unbox neural networks for a geophysical application.
Despite the fact that in the past years CNNs have been extensively employed on seismic applications, a lack of rigorous hyperparameters choice's explanation is still an unsolved issue.
Therefore, we conduct in this work an extensive and detailed set of experiments, bridging together the neural network parameters and their effects for the demultiple task.
While promising, the empirical assessment only represents a subset of the total amount of possible configurations.
Nonetheless, it is sufficient to start deciding which hyperparameters play an important role, and which are considered of secondary importance.

\section{Acknowlegement}
This work was developed in the Fraunhofer Cluster of Excellence Cognitive Internet Technologies.
The authors would like to acknowledge the members of the Fraunhofer ITWM DLSeis consortium (http://dlseis.org) for their financial support.
We are also grateful to Equinor and Volve Licence partners for releasing Volve seismic field data under an Equinor Open Data Licence.


\nocite{*}
\bibliographystyle{plain}
\bibliography{main}

\begin{thebibliography}{10}

\bibitem{berkhout1985seismic}
A.J. Berkhout.
\newblock {\em Seismic Migration: Imaging of Acoustic Energy by Wave Field
  Extrapolation}.
\newblock Number Teil 1 in Developments in solid earth geophysics. Elsevier,
  1985.

\bibitem{berryhill1986deep}
JR~Berryhill and YC~Kim.
\newblock Deep-water peg legs and multiples: Emulation and suppression.
\newblock {\em Geophysics}, 51(12):2177--2184, 1986.

\bibitem{breuer2020deep}
Alexander Breuer, Norman Ettrich, and Peter Habelitz.
\newblock Deep learning in seismic processing: Trim statics and demultiple.
\newblock In {\em SEG Technical Program Expanded Abstracts 2020}, pages
  3199--3203. Society of Exploration Geophysicists, 2020.

\bibitem{bugge2021demonstrating}
Aina~Juell Bugge, Andreas~K Evensen, Jan~Erik Lie, and Espen~H Nilsen.
\newblock Demonstrating multiple attenuation with model-driven processing using
  neural networks.
\newblock {\em The Leading Edge}, 40(11):831--836, 2021.

\bibitem{carvalho1991examples}
FM~Carvalho, Arthur~B Weglein, and Robert~H Stolt.
\newblock Examples of a nonlinear inversion method based on the t matrix of
  scattering theory: application to multiple suppression.
\newblock In {\em SEG Technical Program Expanded Abstracts 1991}, pages
  1319--1322. Society of Exploration Geophysicists, 1991.

\bibitem{choi2019empirical}
Dami Choi, Christopher~J Shallue, Zachary Nado, Jaehoon Lee, Chris~J Maddison,
  and George~E Dahl.
\newblock On empirical comparisons of optimizers for deep learning.
\newblock {\em arXiv preprint arXiv:1910.05446}, 2019.

\bibitem{durall2020watch}
Ricard Durall, Margret Keuper, and Janis Keuper.
\newblock Watch your up-convolution: Cnn based generative deep neural networks
  are failing to reproduce spectral distributions.
\newblock In {\em Proceedings of the IEEE/CVF conference on computer vision and
  pattern recognition}, pages 7890--7899, 2020.

\bibitem{durall2021generative}
Ricard Durall, Valentin Tschannen, Norman Ettrich, and Janis Keuper.
\newblock Generative models for the transfer of knowledge in seismic
  interpretation with deep learning.
\newblock {\em The Leading Edge}, 40(7):534--542, 2021.

\bibitem{gavrikov2022empirical}
Paul Gavrikov and Janis Keuper.
\newblock An empirical investigation of model-to-model distribution shifts in
  trained convolutional filters.
\newblock {\em arXiv preprint arXiv:2201.08465}, 2022.

\bibitem{gisolf2010principles}
D.~Gisolf and E.~Verschuur.
\newblock {\em The principles of quantitative acoustical imaging}.
\newblock EAGE Publ., 2010.

\bibitem{hampson1986inverse}
Dan Hampson.
\newblock Inverse velocity stacking for multiple elimination.
\newblock In {\em SEG Technical Program Expanded Abstracts 1986}, pages
  422--424. Society of Exploration Geophysicists, 1986.

\bibitem{hardt2016train}
Moritz Hardt, Ben Recht, and Yoram Singer.
\newblock Train faster, generalize better: Stability of stochastic gradient
  descent.
\newblock In {\em International conference on machine learning}, pages
  1225--1234. PMLR, 2016.

\bibitem{kingma2014adam}
Diederik~P Kingma and Jimmy Ba.
\newblock Adam: A method for stochastic optimization.
\newblock {\em arXiv preprint arXiv:1412.6980}, 2014.

\bibitem{nair2010rectified}
Vinod Nair and Geoffrey~E Hinton.
\newblock Rectified linear units improve restricted boltzmann machines.
\newblock In {\em Icml}, 2010.

\bibitem{ronneberger2015u}
Olaf Ronneberger, Philipp Fischer, and Thomas Brox.
\newblock U-net: Convolutional networks for biomedical image segmentation.
\newblock In {\em International Conference on Medical image computing and
  computer-assisted intervention}, pages 234--241. Springer, 2015.

\bibitem{sacchi1999fast}
Mauricio~D Sacchi and Milton Porsani.
\newblock Fast high resolution parabolic radon transform.
\newblock In {\em SEG Technical Program Expanded Abstracts 1999}, pages
  1477--1480. Society of Exploration Geophysicists, 1999.

\bibitem{sacchi1995high}
Mauricio~D Sacchi and Tadeusz~J Ulrych.
\newblock High-resolution velocity gathers and offset space reconstruction.
\newblock {\em Geophysics}, 60(4):1169--1177, 1995.

\bibitem{trad2003latest}
Daniel Trad, Tadeusz Ulrych, and Mauricio Sacchi.
\newblock Latest views of the sparse radon transform.
\newblock {\em Geophysics}, 68(1):386--399, 2003.

\bibitem{verschuur1992adaptive}
Dirk~J Verschuur, AJ~Berkhout, and CPA Wapenaar.
\newblock Adaptive surface-related multiple elimination.
\newblock {\em Geophysics}, 57(9):1166--1177, 1992.

\bibitem{verschuur1991surface}
Dirk~Jacob Verschuur.
\newblock Surface-related multiple elimination, an inversion approach.
\newblock {\em Ph. D. Thesis}, 1991.

\bibitem{verschuur2013multiple}
D.J. Verschuur.
\newblock {\em Seismic multiple removal techniques: past, present, future}.
\newblock EAGE Publications bv, 2013.

\bibitem{weglein1999multiple}
Arthur~B Weglein.
\newblock Multiple attenuation: an overview of recent advances and the road
  ahead (1999).
\newblock {\em The Leading Edge}, 18(1):40--44, 1999.

\bibitem{weglein2003inverse}
Arthur~B Weglein, Fernanda~V Ara{\'u}jo, Paulo~M Carvalho, Robert~H Stolt,
  Kenneth~H Matson, Richard~T Coates, Dennis Corrigan, Douglas~J Foster,
  Simon~A Shaw, and Haiyan Zhang.
\newblock Inverse scattering series and seismic exploration.
\newblock {\em Inverse problems}, 19(6):R27, 2003.

\bibitem{weglein1997inverse}
Arthur~B Weglein, Fernanda~Ara{\'u}jo Gasparotto, Paulo~M Carvalho, and
  Robert~H Stolt.
\newblock An inverse-scattering series method for attenuating multiples in
  seismic reflection data.
\newblock {\em Geophysics}, 62(6):1975--1989, 1997.

\bibitem{weglein2011multiple}
Arthur~B Weglein, Shih-Ying Hsu, Paolo Terenghi, Xu~Li, and Robert~H Stolt.
\newblock Multiple attenuation: Recent advances and the road ahead (2011).
\newblock {\em The Leading Edge}, 30(8):864--875, 2011.

\bibitem{wiggins1988attenuation}
J~Wendell Wiggins.
\newblock Attenuation of complex water-bottom multiples by wave-equation-based
  prediction and subtraction.
\newblock {\em Geophysics}, 53(12):1527--1539, 1988.

\bibitem{wilson2017marginal}
Ashia~C Wilson, Rebecca Roelofs, Mitchell Stern, Nati Srebro, and Benjamin
  Recht.
\newblock The marginal value of adaptive gradient methods in machine learning.
\newblock {\em Advances in neural information processing systems}, 30, 2017.

\end{thebibliography}

\end{document}